\documentclass[conference]{IEEEtran}
\IEEEoverridecommandlockouts
\usepackage{cite}
\usepackage{amsmath,amssymb,amsfonts}
\usepackage{algorithmic}
\usepackage{graphicx}
\usepackage{textcomp}
\usepackage{xcolor}
\usepackage{url}
\usepackage{balance}
\def\BibTeX{{\rm B\kern-.05em{\sc i\kern-.025em b}\kern-.08em
    T\kern-.1667em\lower.7ex\hbox{E}\kern-.125emX}}
\begin{document}

\title{Analysis of Political Party Twitter Accounts' Retweeters During Japan's 2017 Election}

\author{\IEEEauthorblockN{Mitsuo Yoshida}
\IEEEauthorblockA{\textit{Toyohashi University of Technology}\\
Toyohashi, Aichi, Japan \\
yoshida@cs.tut.ac.jp}
\and
\IEEEauthorblockN{Fujio Toriumi}
\IEEEauthorblockA{\textit{The University of Tokyo}\\
Bunkyo-ku, Tokyo, Japan \\
tori@sys.t.u-tokyo.ac.jp}
}

\maketitle

\begin{abstract}
In modern election campaigns, political parties utilize social media to advertise their policies and candidates and to communicate to the electorate.
In Japan's latest general election in 2017, the 48th general election for the Lower House, social media, especially Twitter, was actively used.
In this paper, we analyze the users who retweeted tweets of political parties on Twitter during the election.
Our aim is to clarify what kinds of users are diffusing (retweeting) tweets of political parties.
The results indicate that the characteristics of retweeters of the largest ruling party (Liberal Democratic Party of Japan) and the largest opposition party (The Constitutional Democratic Party of Japan) were similar, even though the retweeters did not overlap each other.
We also found that a particular opposition party (Japanese Communist Party) had quite different characteristics from other political parties.
\end{abstract}


\section{Introduction}

A partial amendment of Japan's Public Officers Election Act in 2013 authorized election campaigning using the Internet.
Since then, using social media such as Twitter and Facebook during an election is essential for political parties disseminating information to supporters and gaining new supporters.
Yet there are also reports that the impact of the Internet on elections is limited,
so it is not yet clear what kinds of election strategies in using the Internet are effective in Japan.

In an election using the Internet, political parties expect information diffusion in social media to be strong.
In social media particularly, unlike in conventional media such as newspapers and others, it is possible to spread push-type information diffusion; even for users who are not initially interested in political information, such information is conveyed by spreading behavior, such as the retweeting of other users' tweets.
Therefore, political parties expect that the remarks of their official account will spread to social media as a whole through the cooperation of other users and that the parties' information will be disseminated to more people.

We will analyze users who retweeted the tweets of political party accounts on Twitter during the 48th general election for Japan's Lower House in 2017, aiming to clarify these retweeters' characteristics.
Specifically, in this paper, we analyze what kinds of users retweeted the tweets of each party.
In this analysis, we consider that the quality of information diffusion would vary depending on whether the users who retweeted the official account's tweets were politically interested users or general users.
If there were tweets of official party accounts that were retweeted by users who tweeted few political tweets on a daily basis,
it seems likely that the information had diffused to the general users, so these tweets may have succeeded in appealing to usually unsupportive layers of the populace.
On the other hand, users who tweeted many political tweets on a daily basis were highly likely already to have been supporters of the political party,
so it is considered that tweets with many retweets by such users contributed to appeals to supporters.
Based on this hypothesis, we aim to clarify the differences in Twitter strategy by analyzing users who retweeted tweets of official accounts of each party and analyzing which users did so.

\section{Related Work}

Tumasjan et al. reported that Twitter functions as a discussion forum for politics~\cite{Tumasjan2010},
and there have been many studies of users' access to political information on social media.
Previous studies have shown that users are divided regarding access to political information~\cite{Adamic2005,Hayat2017,Hyun2014,Iyengar2009}.
Such studies mainly cover political information written in the news on social media.
In other words, the studies were not targeted to tweets of political parties and the users who retweeted such tweets.

The number of followers a user has is used as the attention degree of the user.
But the influence of the number of followers in information diffusion is not necessarily large~\cite{Cha2010}, and fraudulent methods are sometimes used to gain followers~\cite{Stringhini2013}.
Even in political communication, the hub of information cannot be determined only by the number of followers~\cite{Bakshy2011}.
On the other hand, few studies have analyzed the tweet strategy of each party based on users' tweet characteristics.

\section{Dataset}

\begin{table*}[tp]
\caption{The major political parties in Japan: ``The Liberal Democratic Party of Japan'' and ``Komeito'' are the ruling parties. We collected the followers and friends on November 10, 2017, and collected others from September 28 to October 23, 2017.}
\centering
\begin{tabular}{l|l|rrrr}
\hline
Party Name & Screen Name & \# of tweets & \# of retweeted & \# of followers & \# of friends \\
\hline
The Liberal Democratic Party of Japan		& @jimin\_koho	& 280	& 110,685	& 134,595	& 322 \\
The Constitutional Democratic Party of Japan	& @CDP2017	& 904	& 506,432	& 191,011	& 77 \\
The Party of Hope						& @kibounotou	& 410	& 21,559	& 13,529	& 152 \\
Komeito							& @komei\_koho	& 289	& 31,072	& 76,743	& 1,259 \\
The Japanese Communist Party			& @jcp\_cc		& 347	& 48,203	& 42,508	& 253 \\
The Japan Innovation Party				& @osaka\_ishin	& 281	& 13,163	& 15,999	& 177 \\
\hline
\end{tabular}
\label{tb:parties}
\end{table*}

In this study, we used Japanese retweets on Twitter collected from September 28\footnote{The Lower House in Japan was dissolved on this day.} to October 23, 2017.
This period encompassed the 48th general election for the Lower House in Japan.
The data were collected using the Twitter Search API\footnote{We constantly searched by the query ``RT lang:ja''.} and amounted to 42,651,648 retweets.

To focus only on data related to politics, we targeted the official accounts of political parties and selected the accounts of the six major parties, shown in TABLE~\ref{tb:parties}.
In the collected retweets, we only used 732,861 retweets (84,043 users) in which the tweets of these political parties were retweeted.
Normally, tweets collected using the Twitter Streaming API and the ``follow'' parameter are used in this type of study.
Since we started this study after the end of this election, we decided to extract the necessary data from the collected retweet data.

Basic information on political party accounts is shown in TABLE~\ref{tb:parties}.
Here, ``\# of tweets'' is the number of tweets (original tweets and retweets) posted by the official account,
and ``\# of retweeted'' the total number of the official account's tweets that were retweeted by users, as followers, and by others.
The designation ``\# of followers'' is the number of users following the official account,
and ``\# of friends'' is the number of users followed by the official account.

Many political party accounts have many followers, but the accounts do not always follow users.
We found that there were political parties who were actively using Twitter and political parties who did not use it much.
It seems that the number of times the tweets were retweeted was greatly affected by the number of followers.
As shown in Table~\ref{tb:parties}, the number of followers of @CDP2017 (The Constitutional Democratic Party of Japan) was large.
As a result, the number of times tweets of @CDP2017 were retweeted increased.

The above data is part of the data used in our previous study~\cite{Yoshida2018}.
In addition, we used tweets of the 84,043 users who retweeted political parties' tweets, gathering tweets that those users posted from April 10, 2007, to November 14, 2017.
As a result, we collected and used 197,267,467 tweets~\footnote{Due to limitations of the Twitter API, it is limited to a maximum of 3,200 tweets per user.}, which included retweets.
In this paper, we call a tweet that is not a retweet the ``original tweet.''

\section{Political Tweets by Users}

\subsection{Learning Political Tweets}

In order to analyze to what extent each user posted political tweets and retweeted political tweets,
we used machine learning to determine whether each tweet was a political tweet.

First, we classified 9,500 original tweets, randomly extracted manually, either as political tweets or as general tweets.
Then, the system learned to classify tweets, using those original tweets as training data, using a support vector machine (SVM), logistic regression, and convolutional neural networks (CNN)~\cite{Kim2014}.
In SVM and logistic regression, the tweets were converted into a bag-of-words normalized by TF-IDF and were learned.
In CNN, words in tweets were converted into word distribution expressions obtained by fastText~\cite{Bojanowski2017} and were learned.

\subsection{Extracting Political Tweets}

For the three methods, the performance was verified using 70\% of the whole as learning data and 30\% as test data.
TABLE~\ref{tb:accuracy} shows the accuracy with the test data.
From this, we confirmed that CNN could extract political tweets with the highest accuracy.
This result shows the same trend as the report by Oliveira et al.~\cite{Oliveira2018} which classified Brazilian tweets.

\begin{table}[tp]
  \centering
  \caption{Accuracy of political tweets.}
  \label{tb:accuracy}
  \begin{tabular}{l|r}
  \hline
method & accuracy \\
  \hline
SVM & 0.671 \\
logistic regression & 0.727 \\
CNN & 0.839 \\
  \hline
  \end{tabular}
\end{table}

\begin{figure}[tp]
  \centering
  \includegraphics[width=0.99\linewidth]{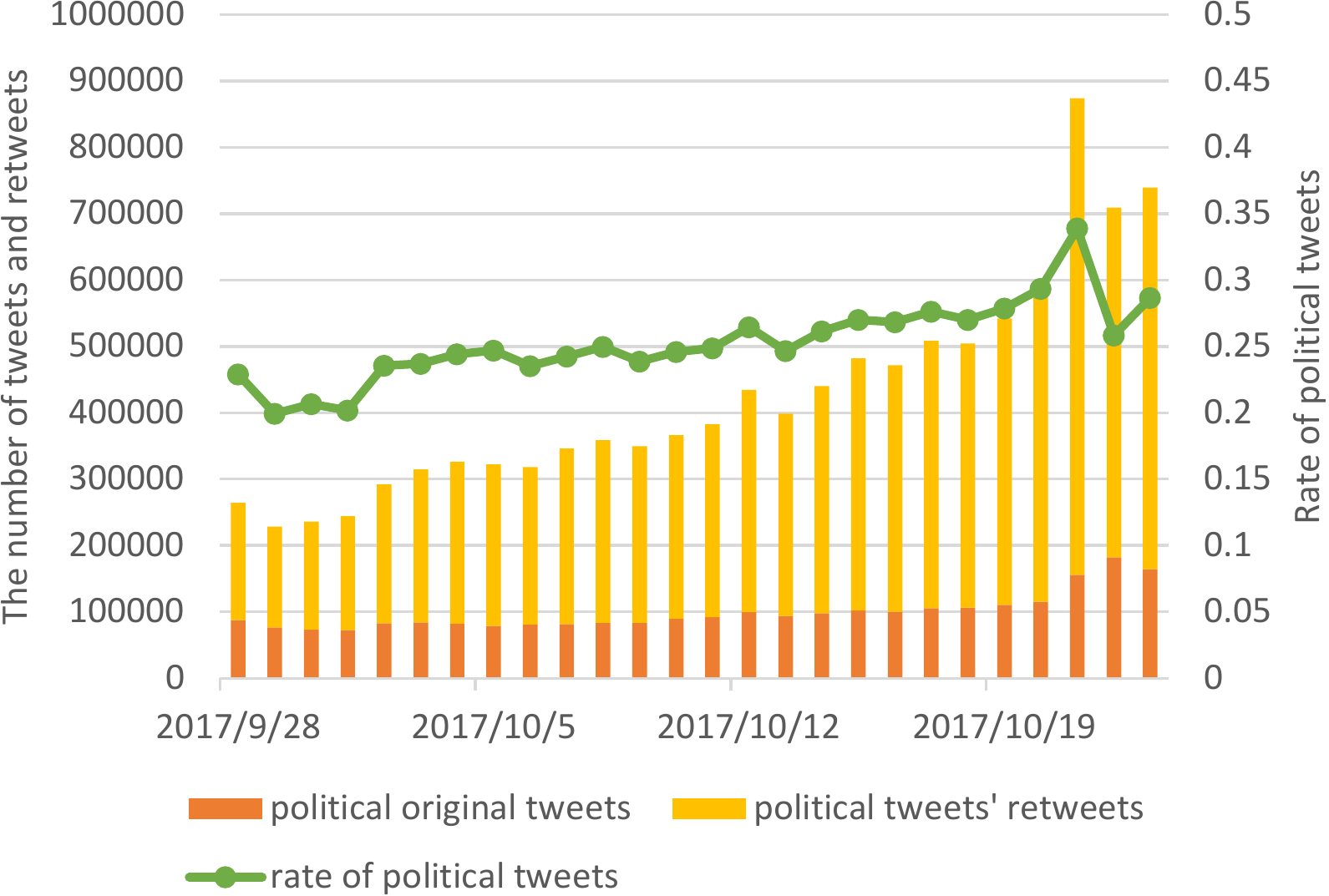}
  \caption{The number of daily political tweets: The left-side y-axis indicates the number of original political tweets and those political tweets' retweets, and the right-side y-axis indicates the ratio of political tweets to the whole. The number of political tweets starts to increase from the day when the Lower House is dissolved, and the peak will come on the eve of the voting day.}
  \label{fig:political_tweets}
\end{figure}

For all tweets collected, we judged whether or not each tweet was a political tweet.
All tweets collected here were the tweets of users who retweeted an official account's tweets.
As a result, 11,635,182 tweets (5.9\% of the number of original tweets and retweets collected) were judged to be political tweets. 
Fig.~\ref{fig:political_tweets} shows the number of political tweets (original tweets and retweets) during the election period (from September 28 to October 23, 2017),
as well as the rate of political tweets in comparison to the whole.
The number of political tweets started to increase from the day when the Lower House was dissolved, and the peak came on the eve of the voting day.
The peak was not on the voting day itself, because under Japanese law, people cannot call for voting during the actual voting time.
From the above, the results are considered to be reasonable.

\section{Analysis of Users Retweeted}

\subsection{Duplication of Users Retweeted}

Users who retweeted the parties official accounts' tweets may have been users who were somehow interested in the policy of the parties.
By analyzing the characteristics of such users, we clarified what kind of user diffused the information of each party account.
Below, the user who retweets the tweets of a party is called a ``retweeter.''

In order to confirm how many users were multiparty retweeters,
we confirmed the rate of users who retweeted tweets of two parties.
The results are shown in TABLE~\ref{tb:duplicate},
which shows the rate of users who retweeted the tweets of the political parties along the top row out of retweeters of the political parties on the left.

\begin{table*}[tp]
  \centering
  \caption{Rate of duplicate retweeters between official accounts: This shows the rate of users who retweeted the tweets of the political parties along the top row out of retweeters of the political parties on the left.}
  \label{tb:duplicate}
  \begin{tabular}{l|rrrrrr}
  \hline
 & @jimin\_koho & @CDP2017 & @kibounotou & @komei\_koho & @jcp\_cc & @osaka\_ishin \\
  \hline
@jimin\_koho &  & 0.101 & 0.032 & 0.104 & 0.035 & 0.033 \\
@CDP2017 & 0.043 &  & 0.021 & 0.022 & 0.148 & 0.005 \\
@kibounotou & 0.187 & 0.285 &  & 0.093 & 0.110 & 0.056 \\
@komei\_koho & 0.222 & 0.110 & 0.034 &  & 0.042 & 0.023 \\
@jcp\_cc & 0.070 & 0.708 & 0.038 & 0.040 &  & 0.009 \\
@osaka\_ishin & 0.380 & 0.131 & 0.111 & 0.126 & 0.050 & \\
  \hline
\end{tabular}
\end{table*}

Many users retweeted the tweets of a single party only.
But exceptional parties with a high rate of retweeters retweeting other parties were as follows:
\begin{enumerate}
  \item Retweeters of @kibounotou and @jcp\_cc (The Japanese Communist Party) retweeted tweets of @CDP2017 (The Constitutional Democratic Party of Japan).
  \item Retweeters of @komei\_koho (The Liberal Democratic Party of Japan) and @osaka\_ishin (The Japan Innovation Party) retweeted tweets of @jimin\_koho (The Liberal Democratic Party of Japan).
\end{enumerate}
Considering the process of establishing @CDP2017 (@CDP2017 was founded independently of @kibounotou.),
it would be reasonable for retweeters of @kibounotou to retweet the tweets of @CDP2017.
Also, since @jimin\_koho and @komei\_koho were the ruling parties in the coalition government,
it would be reasonable for retweeters of @jimin\_koho and @komei\_koho to retweet the tweets of both parties.
On the other hand, 71\% of @jcp\_cc retweeters retweeted the tweets of @CDP2017, but only a few retweeters of @CDP2017 also retweeted the tweets of @jcp\_cc.
This indicates that empathy concerning these political parties or their policies did not extend in both directions.

\subsection{Clustering of Users Using Political Tweets}

We conducted clustering based on the rate of political tweets of the retweeters of each party.
By doing this, we analyzed the characteristics of the retweeters of each party.

First, the following was calculated for each retweeter based on the tweets during the election period:
\begin{itemize}
  \item the number of original tweets ($tw$);
  \item the number of retweets ($rt$);
  \item the number of original political tweets ($ptw$); and
  \item the number of people retweeting political retweets ($prt$).
\end{itemize}
Then we calculated the following as features:
\begin{itemize}
  \item the rate of retweets ($rt/tw$);
  \item the rate of original political tweets ($ptw/(tw-rt)$); and
  \item the rate of retweeting of political retweets ($prt/rt$).
\end{itemize}
Based on these features, retweeters were clustered by K-Means clustering ($k=7$).

The obtained clusters are shown in TABLE~\ref{tb:clusters};
the characteristic cluster C1 has few retweets and few original political tweets, but the retweets contain 83.5\% political content.
C2 has few retweets and few original political tweets.
C4 has many retweets, but the numbers of both original political tweets and political retweets are small.
C6 has many retweets, and there are many original political tweets among these retweeters' tweets.

\begin{table*}[tp]
  \centering
  \begin{tabular}{cc}

    \begin{minipage}[t]{0.35\hsize}
  \centering
  \caption{User clusters based on political tweets.}
  \label{tb:clusters}
  \scalebox{0.9}{
  \begin{tabular}{l|rrr|r}
  \hline
\# & $rt/tw$ & $ptw/(tw-rt)$ & $prt/rt$ & \# of users \\
  \hline
C1 & 0.304 & 0.402 & 0.835 & 1,991 \\
C2 & 0.332 & 0.110 & 0.167 & 7,662 \\
C3 & 0.421 & 0.576 & 0.391 & 4,690 \\
C4 & 0.825 & 0.094 & 0.159 & 8,772 \\
C5 & 0.844 & 0.425 & 0.391 & 9,081 \\
C6 & 0.865 & 0.809 & 0.498 & 6,204 \\
C7 & 0.951 & 0.032 & 0.564 & 6,617 \\
  \hline
  \end{tabular}
  }
    \end{minipage}

    \begin{minipage}[t]{0.65\hsize}
  \centering
  \caption{Similarity of user clusters based on political tweets.}
  \label{tb:correlation}
  \scalebox{0.9}{
  \begin{tabular}{l|rrrrr}
  \hline
 & @CDP2017 & @kibounotou & @komei\_koho & @jcp\_cc & @osaka\_ishin \\
  \hline
@jimin\_koho & 0.816 & 0.731 & 0.539 & 0.470 & 0.827 \\
@CDP2017 &  & 0.454 & 0.801 & 0.471 & 0.557 \\
@kibounotou &  &  & 0.455 & 0.137 & 0.624 \\
@komei\_koho &  &  &  & 0.004 & 0.097 \\
@jcp\_cc &  &  &  &  & 0.731\\
  \hline
  \end{tabular}
  }
  \bigskip
  \bigskip
    \end{minipage}

  \end{tabular}
\end{table*}

\subsection{Cluster of Retweeters in Each Party}

\begin{figure}[tp]
  \centering
  \includegraphics[width=0.99\linewidth]{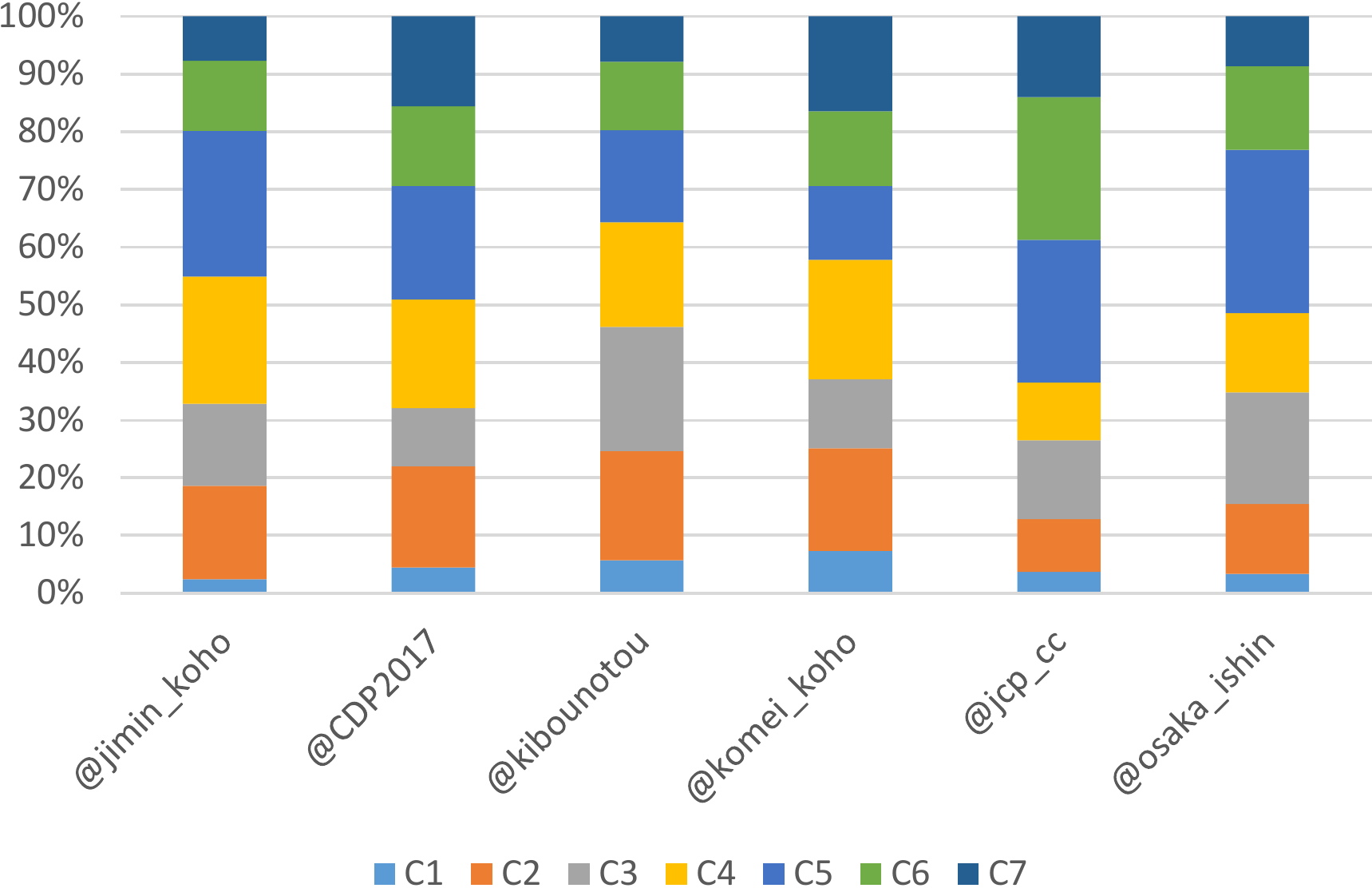}
  \caption{Rate of retweeters' belonging to clusters by political party. The y-axis indicates the rate of a cluster among retweeters of a party. Many parties have similar tendencies.}
  \label{fig:clusters}
\end{figure}

We analyzed which clusters the retweeters of each party belonged to.
Fig.~\ref{fig:clusters} shows the results for how many retweeters of each cluster existed for each party.
Many parties had similar tendencies.

In order to clarify the characteristics of each party,
we calculated the correlations of the rate of retweeters of each party.
This correlation was calculated based on the values in Fig.~\ref{fig:clusters}.
If the correlation was high, this meant that the retweeters belonged to a similar user group,
while if the correlation was low, this meant that the characteristics of the retweeters were different.
The results are shown in TABLE~\ref{tb:correlation}.
The correlation coefficient (0.82) between @jimin\_koho (The Liberal Democratic Party of Japan) and @CDP2017 (The Constitutional Democratic Party of Japan) was extremely high.
This meant that the retweeters of these two parties were similar,
and we found that neither group had any outstanding characteristics.
The retweeters for the accounts @jimin\_koho and @osaka\_ishin (The Japan Innovation Party) were also similar, and those of @CDP2017 and @komei\_koho (Komeito) were similar as well.
However, these pairs were composed of completely different users, because, as shown in TABLE~\ref{tb:duplicate}, there was little duplication of retweeters.
However, the characteristics of the retweeters belonging to each group were similar.
On the other hand, the retweeters of @jcp\_cc (The Japanese Communist Party) were similar only to the retweeters of @osaka\_ishin,
and we found that these retweeters had quite different characteristics from those of other political parties.

\section{Conclusion}

We analyzed the users (retweeters) who retweeted tweets of political party accounts on Twitter during the 48th general election for Japan's Lower House in 2017.
First, we classified retweeters' tweets as political tweets or not, and we clustered retweeters based on the rate of their political tweets.
Then analyzed the rate of users belonging to each cluster in relation to the number of retweeters of each party.
The results indicated that the characteristics of retweeters of @jimin\_koho (The Liberal Democratic Party of Japan) and @CDP2017 (The Constitutional Democratic Party of Japan) were similar,
even though the retweeters did not overlap each other.
We also found that retweeters of @jcp\_cc (The Japanese Communist Party) were similar only to retweeters of @osaka\_ishin (The Japan Innovation Party),
and we found that these retweeters had quite different characteristics from those of other political parties.

\bibliographystyle{IEEEtran}
\bibliography{IEEEabrv,references}

\end{document}